\documentclass[preprint,preprintnumbers,amsmath,amssymb]{revtex4}
\usepackage{graphicx}

\begin{document}
\title{Effects of current on nanoscale ring-shaped magnetic tunnel junctions}

\author{Hong-Xiang Wei$^1$, Jiexuan He$^2$, Zhen-Chao Wen$^1$, Xiu-Feng Han$^{1*}$, Wen-Shan Zhan$^1$, and Shufeng Zhang$^{2*}$}
\affiliation{$^1$State Key Laboratory of Magnetism, Beijing
National Laboratory for Condensed Matter Physics, Institute of
Physics, Chinese Academy of Science, Beijing 100080, China \\
$^2$Department of Physics and Astronomy, University of Missouri,
Columbia, MO 65211}

\thanks{Corresponding authors: xfhan@aphy.iphy.ac.cn and ZhangShu@missouri.edu}

\begin{abstract}
We report the observation and micromagnetic analysis of
current-driven magnetization switching in nanoscale ring-shaped
magnetic tunnel junctions. When the electric current density exceeds
a critical value of the order of $6\times 10^{6}$A/cm$^2$, the
magnetization of the two magnetic rings can be switched back and
forth between parallel and antiparallel onion states. Theoretical
analysis and micromagnetic simulation show that the dominant
mechanism for the observed current-driven switching is the spin
torque rather than the current-induced circular Oersted field.
\end{abstract}

\maketitle

\section{Introduction}

In the experiments that demonstrated current-driven magnetization
switching for spin valve pillars \cite{Ralph} or magnetic tunnel
junctions (MTJs) \cite{Huai}, magnetic elements have been patterned
into elliptic or rectangular shapes. Undesired properties such as a
large shape anisotropy and a strong stray field would place a severe
limitation on these magnetic elements for ultra-high density memory
devices. The ring-shaped spin valve or MTJ eliminates the stray
field and enhances the thermal stability since the magnetization may
ideally form a vortex structure free of magnetic poles
\cite{Zhu,Chien,Chu}. Recently, the magnetic rings made of a
single magnetic layer \cite{Vaz,LiS,Rotman} and of spin valves
\cite{Ross,Ross1,Otani,Zhu2,Chen} have been achieved. For the current in
the plane of the layers (CIP), the measurement of the CIP resistance
of the spin valve probes the domain structure of the free layer \cite{Ross}.
For the current perpendicular to the plane of the layers (CPP), the
magnetization can be switched by the current \cite{Otani,Zhu2,Chen}.
In these studies, the size of the rings is ranged from sub-micrometers to
micrometers and the current-induced Oersted field is the dominant
mechanism in switching. To achieve the switching by the spin transfer
torque \cite{Slonczewski}, it is necessary to make the ring size of the order
of 100nm or less. We had
previously reported ring-shaped magnetic tunnel junctions \cite{Wen}
where the fixed layer is pinned by an antiferromagnetic layer.

In the present work, we have successfully made a series of the
ring-shaped MTJ structures with various ring diameters and without an
antiferromagnetic layer; this simpler layer structure makes the
analysis more straightforward. Our micromagnetic simulation shows
that the observed magnetization switching is originated from spin
transfer torques and the current-induced circular Oersted field
plays very minor roles in switching. Although the desired vortex states
have not been identified in our experiments and simulation, the definitive
verification of the switching between onion states (described in detail later)
by the spin torque is encouraging for further exploring the geometrically
controlled magnetic states.

\section{Experimental procedures}

Multilayered MTJ films with the
hard-ferromagnetic/insulator/soft-ferromagnetic layer structure of
Ta(5)/Cu(20)/Ta(5)/Co$_{50}$Fe$_{50}$(2.5)/Al(0.7)-oxide/Co$_{60}$Fe$_{20}$B$_{20}$(2.5)/Ta(3)/Ru(5)
(thickness unit: nm) were deposited on the Si(100)/SiO$_2$ substrate
using an ULVAC TMR R \& D Magnetron Sputtering System (MPS-4000-HC7)
with a base pressure of $1\times 10^{-6}$ Pa. The Al-oxide barrier
was fabricated by inductively coupled plasma (ICP) oxidizing 0.6 nm
Al-layer with an oxidation time of 10 s in a mixture of oxygen and
argon at a pressure of 1.0 Pa in a separate chamber. Two electrodes
situated above and below the NR-MTJ were patterned by ultraviolet
optical lithography (UVL) combined with Ar-ion beam
milling \cite{Zeng,Wei}. The active ring-shaped junction area was
patterned by electron beam lithography (EBL) using a Raith 150
scanning electron microscope and reactive ion etching (RIE). The
nanoring MTJ pillar including the top resist was then buried by
SiO$_2$ deposition. Finally, the resist and SiO$_2$ on the top of a
nanoring were removed using a lift-off process before the top
electrode was patterned in the perpendicular direction.

The transport measurement was conducted via a standard four probe
method. The tunnel resistance was measured by applying a small current
($10 \mu A$) so that the magnetic state of the ring is not disturbed.

\section{Experimental Results and Micromagnetic Analysis}

Figure 1(a-b) show the SEM images of an array of ring-shaped MTJs
and of a single MTJ whose inner and outer diameters are 50nm and 100
nm, respectively. The magnetization state of a deep submicron-sized
single ring has two possible stable configurations at the zero
external magnetic field (remanent state):  onion (O) and vortex (V)
as shown in Fig.~1(d). The onion states are metastable states; they
can be easily formed via the application of an in-plane magnetic
field. Each onion state has a pair of domain walls; in the presence
of a large uniaxial anisotropy, two domain walls are pinned around
the easy axis. If the anisotropy is small, the magnetostatic
interaction between two domain walls tends to attract each other,
forming a slightly asymmetric onion state (not shown). The vortex
state is the lowest energy state and is more stable than the O
state. However, there is no simple experimental method available
to produce the V state. Our micromagnetic simulation \cite{simulation} shows
that the transition from the onion state to the vortex state does
not occur when one just sweeps the in-plane magnetic field from a
large positive value to a negative one for the samples with only
2.5nm layer thickness. Instead, the two domain walls of the
onion state rotate along the circumference until the polarization of the
onion state is reversed. We will discuss the generation of the V
state in the end of the paper.

\subsection{Magnetic field driven switching}

Figure 2 shows the tunnel resistance when we apply an in-plane
magnetic field along the easy axis of the ring. At a large field,
magnetization of both rings are in the
parallel onion state and thus the resistance is lowest. When the
field is reduced, the stray field of the domain walls causes the
onion states of the upper and lower rings to rotate oppositely in
order to reduce the magnetostatic energy. Since the anisotropy of
the two layers is small, each onion
state will oppositely rotate 90$^0$ to form an antiparallel
configuration and the tunnel resistance becomes maximum. Further
reversing the magnetic field causes the onion states of the two
layers to become parallel again.

To reproduce the qualitative
features of the above experimental data, we perform a numerical
simulation by taking the following plausible assumptions. First,
several jumps shown in Fig.~2a indicate the existence of pinning
potentials and we model the pinning by simply including an in-plane
four-fold anisotropic field. Second, an interlayer coupling between
two rings is likely to exist for the ultra-thin $Al_2O_3$ insulator
barrier. Thus, we have added a 60 (Oe) ferromagnetic coupling field,
which is consistent with our previous experiments on similar
junctions having elliptical shapes \cite{Han}. Third, the tunnel
conductance $G(H)$ is calculated via the standard formulation of
tunnel magnetoresistance given below
\begin{equation}
G(H)= G_p - \frac{G_p-G_a}{2A} \int dxdy [1-\cos\theta (x,y)]
\end{equation}
where $G_p$ ($G_a$) are the conductance for the two magnetic layers
in parallel (antiparallel), $A$ is the area of the ring, $\theta
(x,y)$ is the angle between the local magnetization vectors $
\vec{M}_t(x,y)$ and $\vec{M}_b (x,y)$ of the top and bottom rings,
and $(x,y)$ denotes the in-plane coordinates. Once the magnetization
states of the top and bottom rings are obtained via micromagnetic
simulation, the conduction $G(H)$ or the tunnel magnetoresistance
$R(H)=G^{-1}(H)$ can be readily derived from Eq.~(1). As seen from
Fig.~2(b), the simulated results reproduce the main shape of the
experimental R-H curves.

We note that there are several subtle features
that are not reproduced by the simulation. First, the gradual
increase of the resistance at low field observed in the experiment
indicates the gradual rotation of the domain walls; the origin is
likely from the competition between magnetostatic interactions of
the two rings and the pinning potentials \cite{Hayward}. Second,
there are small sudden jumps in the resistance at low fields between
the parallel resistance and the
onset of the gradual increase to the anti-parallel resistance.
The reason might be due to the misalignment among
easy axes of two layers and the direction of the magnetic field.
If we consider that the easy axes of the two layers make an angle $\theta$, a
large field would make the onion states of the two layers in parallel. As
the field reduces to a smaller value, a transition occurs where the two
onion states tend to return to their respective easy axis, i.e., forming
approximately an angle $\theta$. Note that this irreversible jump depends on
the location of the pinning potential. We show in Fig.~3, the R-H
characteristic of a similar ring
where the major and minor hysteresis jumps occur in
several magnetic fields. Interestingly, the inserted minor loop contains
only one small
jump. These uncontrolled subtle feature indicates the imperfection of the
ring stucture.

\subsection{Current-driven switching}

Next we show the tunnel resistance by sweeping the current at
zero magnetic field. The data were recorded as follows. Before each
resistance measurement, a current pulse whose amplitude $I$ and
width $200ns$ was applied. The resistance is then measured by using
a low readout current of $10 \mu A$ that will not disturb the
magnetic state formed after each current pulse. By repeating the
above process for an increasing or decreasing amplitude of $I$, we
obtained the full R-I loop in Fig.~(4). One notices that the two
values of the resistance are very close to the values in Fig.~2(a),
indicating that the magnetization of the two layers are two onions
states with parallel (low resistance) and antiparallel (high
resistance) configurations. The critical switching current from
antiparallel (parallel) to parallel (antiparallel) states is about
1.1 m$A$, which corresponds to the current density about $6\times
10^{6} A/cm^2$.

Two competing mechanisms are responsible for the switching of the
magnetization by the current. First, the current induces a circular
magnetic field that may affect the magnetization states. However,
our data suggest that the current-induced magnetic field cannot be
the dominant mechanism for the switching: 1) For the current
$I=1$mA, the maximum Oersted field at the outer boundary of an
100nm-diameter infinite long cylinder is less than 40 (Oe); this
Oersted field is clearly an overestimation since the field would be
smaller for our ring structure. Simulation indicates that one needs
at least several times larger current density to reverse the
polarity of the onion state. More specifically, we find from
micromagnetic simulation that without the spin torque the circular
Oersted field is able to rotate the domain walls of the onion state
by only a few degrees. 2) The measured critical current density,
shown in Fig.~(4a), is approximately same for different sizes of
rings (not shown) while the Oersted field driven reversal \cite{Chen}
would be strongly size dependent. Thus we propose that the spin transfer
torque is the dominant factor for the observed switching.

We have performed micromagnetic simulation by explicitly taking into
account both the spin torque and the induced Oersted field. We model the
spin torque on ${\bf m}_i$ via $a_J {\bf m}_i\times ({\bf m}_i
\times {\bf m}_j)$ \cite{ZLi} where $a_J$ is proportional to the current.
Figure 4(b)-(d) show the simulation results.
We note that both the observed (Fig4a) and simulated (Fig4b)
R-I loops does not show distinct steps (see Fig.~2); this indicates that the
current driven switching is relatively insensitive to local pinnings
compared to the field driven switching. We attribute this feature to
the spin transfer torque: the switching current density $I_c$ is
proportional to $H_K+2\pi M_s$ where $H_K$ is the anisotropy field
and $M_s$ is the saturation magnetization \cite{Slonczewski}.
Since $2\pi M_s$ is
much larger than $H_K$ and pinning potentials, the critical current
is thus not sensitive to the local defects.
We also notice that the simulation shows a gradual change in resistance
just before switching near the parallel (anti-parallel) resistance at
positive (negative) currents, while the experimental
switching between parallel and anti-parallel is abrupt. This discrepancy
is due to an artifact that the simulation was carried out with a constant
current density. When the current is near
the switching current, one of the layers is in a stable precessional state
(this stable precessional state is the signature of the spin torque), see
Fig.~3(c). Experimentally, the
resistance was measured after the pulsed current was switched off. Therefore,
the measured states returned to the parallel onion states. If we
turn off the current after certain time in simulation, the loop will
have no gradual change because
the precessional states will return to either the parallel or
antiparallel states, see Fig.~4(e).

\subsection{Discussions and summary}

The quantitative analysis of the spin torque effect in MTJ involves
a number of key parameters. The critical current at finite
temperature is given by \cite{Koch,Li}
\begin{equation}
I_c = I_{c0} \left( 1-\frac{k_BT^*}{E_b} \ln(t_p f_0) \right)
\end{equation}
where $I_{c0}$ is the intrinsic (zero temperature), $T^*$ is the
temperature, $E_b$ is the energy barrier separating antiparallel and
parallel onion states, $t_p =200$ns is the pulse width of the
current and $f_0 \approx 10^9$/s is the attempt frequency. It is
expected that the temperature will be significantly higher than the
room temperature when the bias voltage is of the order of 1 V
\cite{Toshiba}. The energy barrier depends on the detail of the
magnetostatic interaction as well as defects of the ring structure
\cite{Martinez,He}.
Furthermore, the recently proposed perpendicular spin torque also
contributes the energy barrier \cite{SL,Theodonis}. The uncertainty
of these parameters makes the quantitative comparison very
difficult, if not at all impossible. We show in Fig.~(5) the critical
current as a function of the temperature.
The significant reduction of the critical current at high temperatures
implies the importance of the thermally assisted switching.

While our initial motivation for the study of ring structure is to
create and switch vortex states which are robust against thermal
fluctuation, we are unable to create such stable vortex states in
the present experiments. A plausible interpretation is that the
vortex states are topologically different from the onion states and
one needs to annihilate two magnetic domains of the onion state to
create a vortex state. For our ultra-thin rings, one would require a
large out-of-plane rotation to convert the onion to the vortex
states and thus it is energetically prohibited. One possible way to
initialize a vortex state is to utilize a circular magnetic field
generated by the current, but the required current density is too
high. Other methods, for example, applying a large out-of-plane
field along with a moderate current, might be more suitable for the
initial formation of the vortex states. It is an experimental
challenging problem to manipulate well-controlled vortex states in
ultra thin rings. We defer this study in the future.

In summary, we demonstrate for the first time that magnetization of
the ring-shaped MTJ can be switched by a current. The current
density of the order of $6\times 10^{6} A/cm^2$ is sufficient to
switch one onion state to another. This work opens a possibility for
creating high-density magnetic elements with enhanced thermal
stability and reduced power consumption.

This research was supported by the State Key Project of Fundamental
Research of Ministry of Science and Technology (MOST), Chinese
Academy of Sciences (CAS), and National Natural Science Foundation
(NSFC), China. X. F. Han gratefully thanks the partial support of
Outstanding Young Researcher Foundation (Grant No.50325104 and
50528101), NSFC project(Grant No.10574156), Wang Kuan-Cheng
Foundation and Micro-fabrication Center of IOP from CAS. S. Zhang
acknowledges the support from NSF (DMR-0704182) and DOE
(DE-FG02-06ER46307) of US.

\pagebreak

\pagebreak
\bigskip
\noindent {Figure Caption}

\bigskip

\noindent{FIG.1} The SEM images of the array {\bf a)} and single
{\bf b)} magnetic tunnel junction rings. Schematically shown are the
layers of the ring {\bf c)} and two stable magnetization patterns
{\bf d)} of one magnetic layer at zero magnetic field (remanent
states): Vortex (V) and Onion (O) states. Note that there are other
equivalent states corresponding to reversing the polarity of the
magnetization of the above two states.

\bigskip

\noindent{FIG.2} The room-temperature tunnel resistance as a
function of the magnetic field for a ring-shaped MTJ whose outer
diameter is 100nm. {\bf b)} Calculated normalized tunnel resistance
$\delta R/\delta R_{max}$ where $\delta R = R(H)-R_p$ and $\delta
R_{max} = R_a-R_p $, and $R_p$ and $R_a$ are the resistances for the
parallel and the antiparallel aligned onion states of the top and
bottom rings. The parameters in micromagnetic simulation are
specified below: the uniaxial anisotropy of FeCo is 50 (Oe) and of
FeCoB is 0 (Oe), the saturation magnetization of FeCo and FeCoB are
1130 emu/cc and 1000 emu/cc, and the interlayer coupling field is 60
(Oe). In addition, an in-plane four-fold anisotropic field of 90
(Oe) is applied to mimic the pinning potentials.

\bigskip

\noindent{FIG.3}The room-temperature tunnel resistance as a function
of the magnetic field for a ring-shaped MTJ whose outer diameter is
200nm. Insert: the minor hysteresis loop.

\bigskip

\noindent{FIG.4}The tunnel resistance as a function of the amplitude
of the 200ns pulsed current density at zero magnetic field. {\bf a)}
Experimental results at room temperature. Simulation results of the
R-I loop {\bf b)}, the precessional states near the critical current
density {\bf c)} and the switched state for the current exceeding
the critical vale. The same set of parameters in Fig.~2 in addition
to the damping parameters of 0.01 and 0.015 for the top and bottom
magnetic rings are used for the simulation. {\bf e)} The simulated
R-I curve after the pulsed current is turned off.

\bigskip

\noindent{FIG.5}The temperature dependence of the dc critical
current density for a ring-shaped MTJ whose outer diameter is 100nm.

\newpage

\begin{figure}
\centering
\includegraphics[width=14cm]{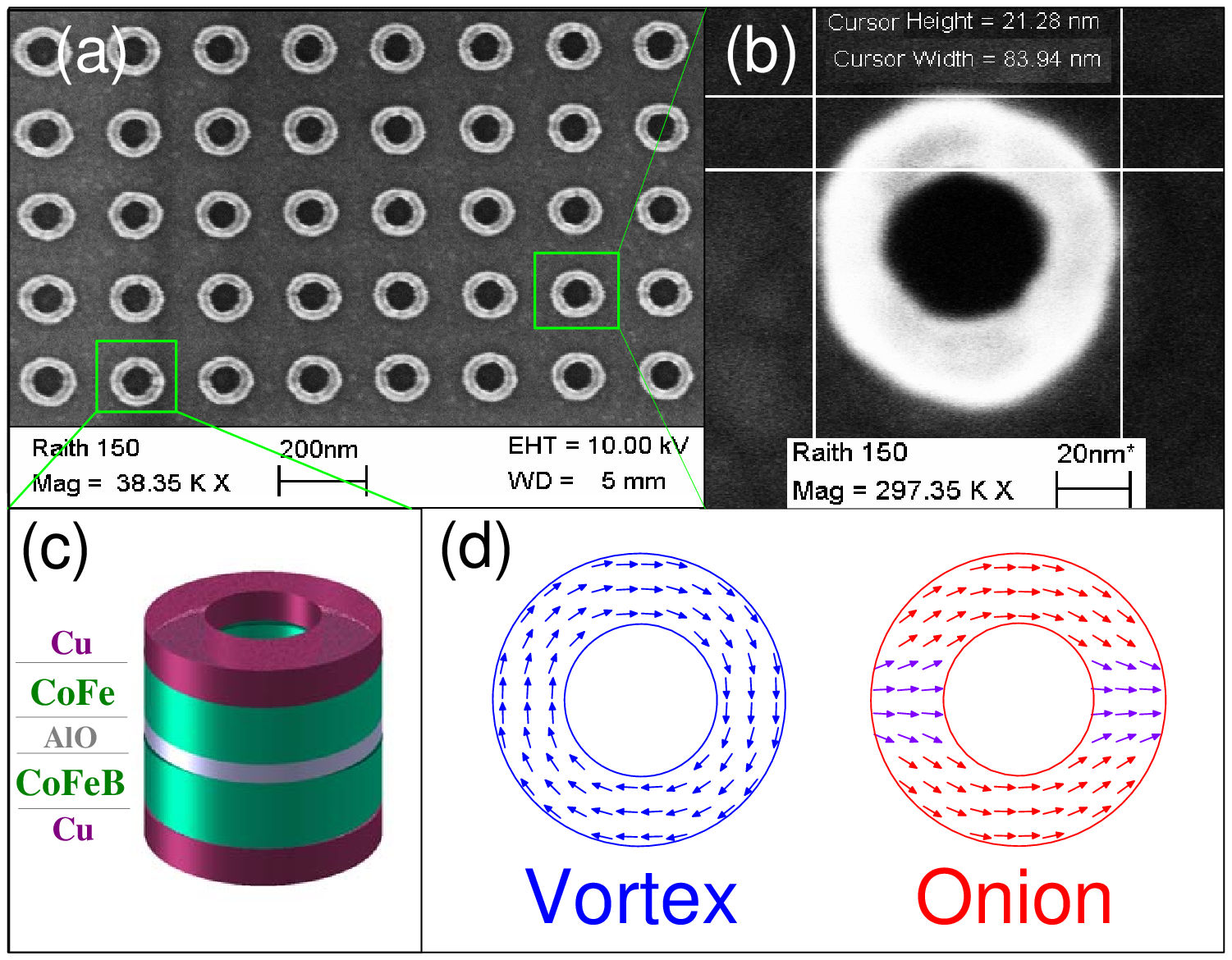}
\caption{}
\end{figure}

\bigskip
\bigskip
\bigskip

\begin{figure}
\centering
\includegraphics[width=14cm]{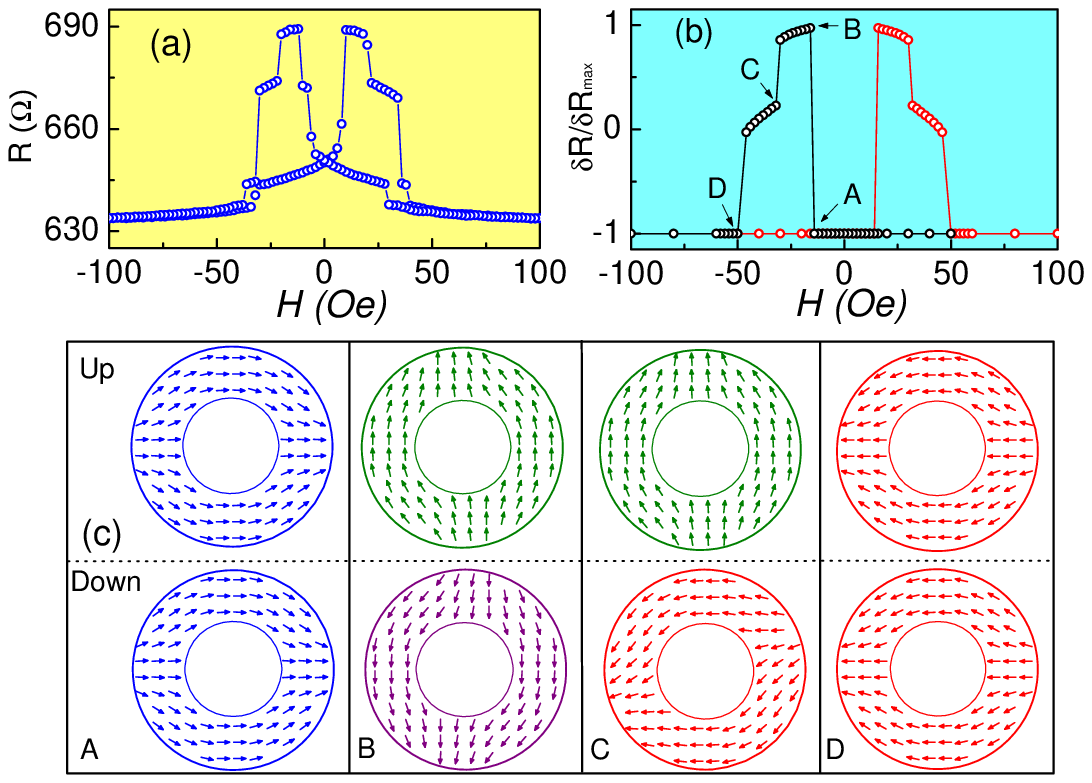}
\caption{}
\end{figure}

\pagebreak
\newpage
\bigskip

\begin{figure}
\centering
\includegraphics[width=14.5cm]{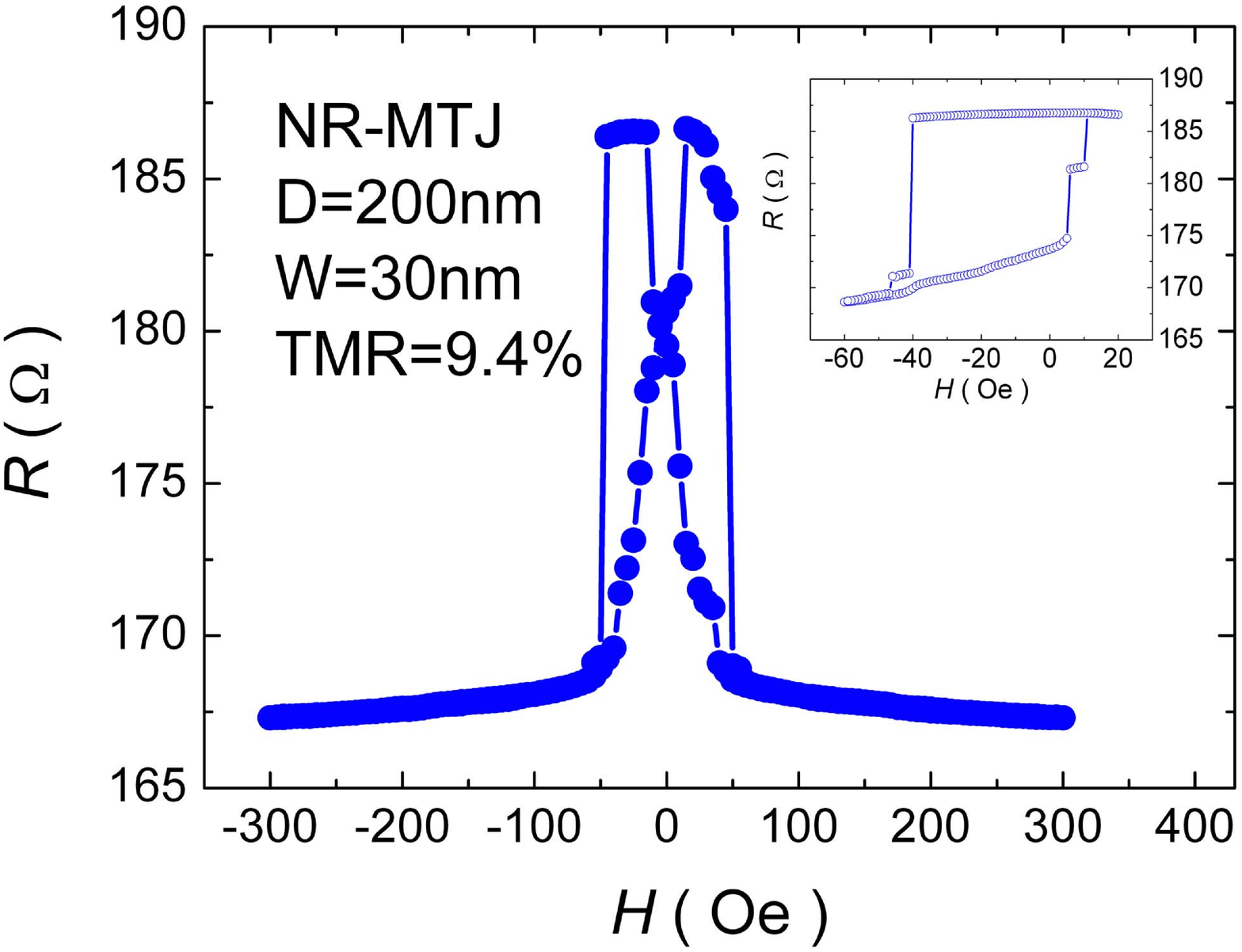}
\caption{}
\end{figure}

\pagebreak
\newpage
\bigskip

\begin{figure}
\centering
\includegraphics[width=14.5cm]{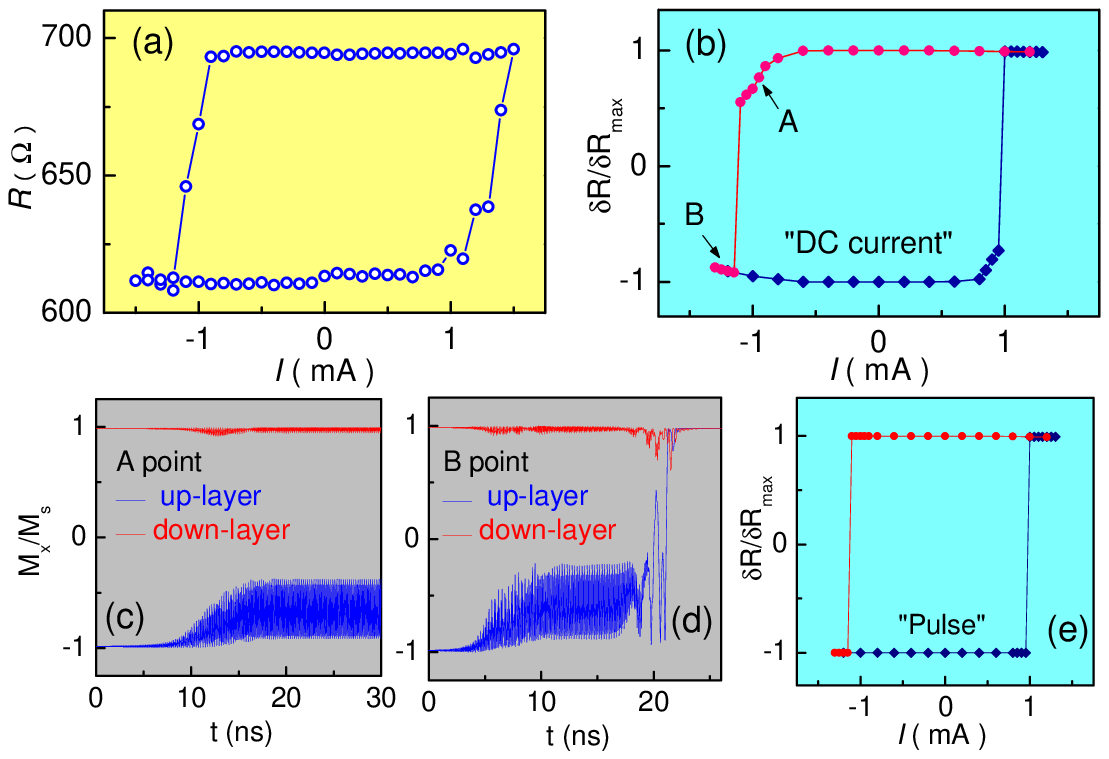}
\caption{}
\end{figure}

\pagebreak
\newpage
\bigskip

\begin{figure}
\centering
\includegraphics[width=14cm]{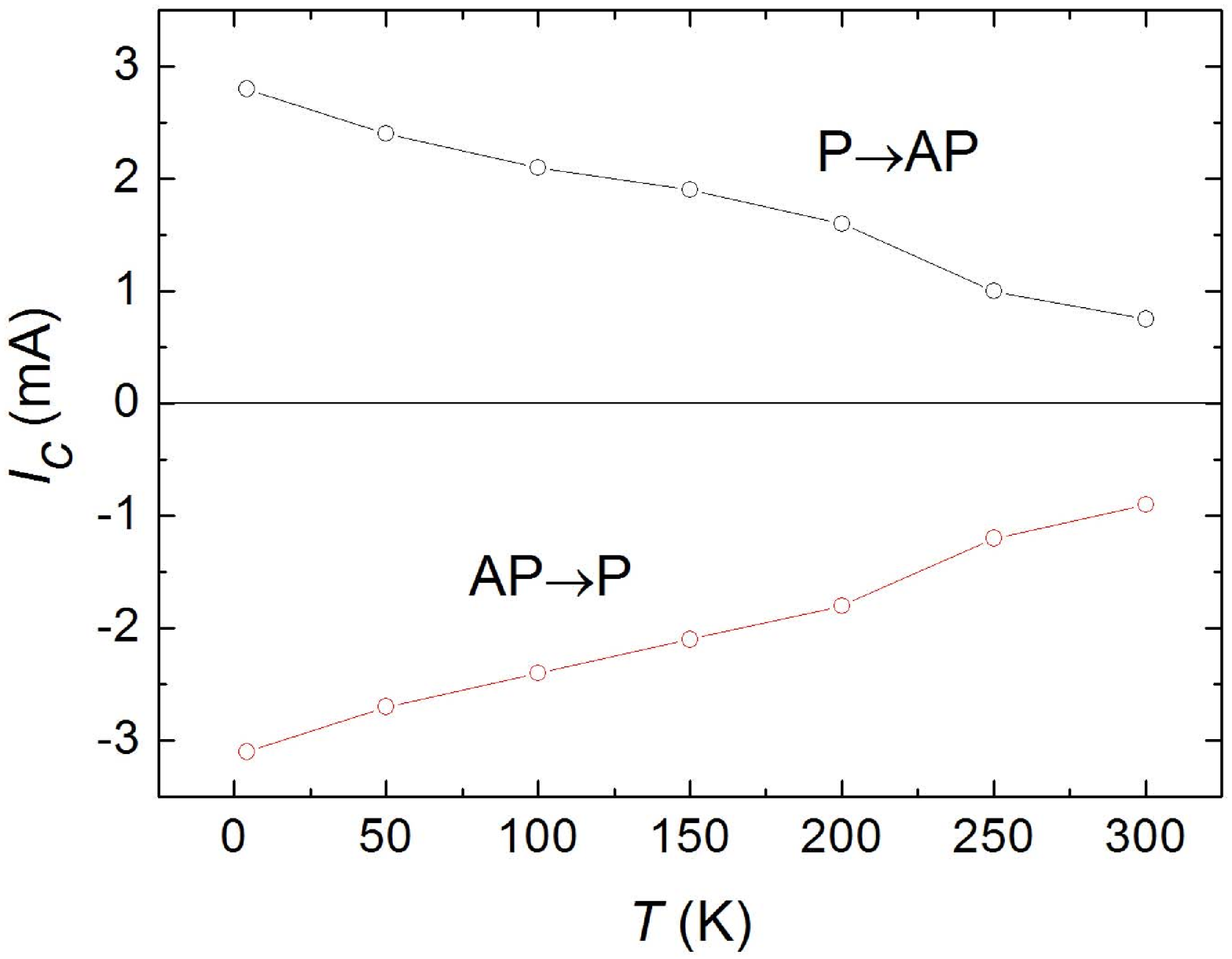}
\caption{}
\end{figure}


\begin{references}
\bibitem{simulation}
Zero-temperature micromagnetic simulation is carried out by choosing
the mesh size $5\times 5\times 0.8$ nm$^3$ so that there are total
seven meshes in the thickness dimension (the center layer is
$Al_2O_3$ that does not have the magnetic moment) throughout the
paper. A special care is taken for the edges of the rings; see M. J.
Donahue, and R. D. McMichael, IEEE Trans. Magn., {\bf 43}, 2878-2880
(2006).
\end{references}
\end{document}